\documentclass{aastex}
\usepackage{spr-astr-addons}
\begin{document}
\title{
The Particle Production at the Event Horizon of a Black Hole
as Gravitational Fowler-Nordheim Emission
in Uniformly Accelerated Frame, in The Non-Relativistic Scenario  
}
\author{Sanchari De$^a$} \and \author{Sutapa Ghosh$^b$} 
\and \author{Somenath Chakrabarty$^{a\dagger}$}
\affil{
$^a$Department of Physics, Visva-Bharati, Santiniketan-731235,
India \\
$^b$Department of Physics, Barasat Govt. College, 
Barasat 700124, India \\
$^\dagger$E-mail: somenath.chakrabarty@visva-bharati.ac.in}
\begin{abstract}
In the conventional scenario, the  Hawking radiation is believed to
be a tunneling process at the event horizon of the black hole. In the
quantum field theoretic approach the Schwinger's mechanism is generally
used to give an explanation of this  tunneling process. It is the decay of
quantum vacuum into particle anti-particle pairs near the black hole
surface. However, in a reference frame undergoing a uniform accelerated 
motion in an
otherwise flat Minkowski space-time geometry, in the non-relativistic
approximation, the particle production near the event horizon of a
black hole may be treated as a kind of
Fowler-Nordheim field emission, which is  the typical electron emission 
process from a metal surface under the action of an external
electrostatic field. This type of emission from metal surface is
allowed even at extremely low temperature. It has been noticed that in
one-dimensional scenario, the Schr$\ddot{\rm{o}}$dinger equation
satisfied by the created particle (anti-particle) near the event
horizon, can be reduced to a differential form which is exactly identical 
with that obeyed by an
electron immediately after the emission from the metal surface under the 
action of a strong
electrostatic field. The mechanism of particle production near the
event horizon of a black hole  is therefore
identified with Schwinger process in relativistic quantum 
field theory, whereas in the non-relativistic scenario it may be
interpreted as
Fowler-Nordheim emission process, when observed from a uniformly
accelerated frame.
\end{abstract}
\maketitle
\section{Introduction}
During the last few decades a lot of work have been reported on the 
identical nature of Schwinger mechanism of pair production in presence
of strong electric field
\cite{SCHW} (see also \cite{LCB,SPK} and references therein) and the
Hawking radiation \cite{HW1,HW2} (see also \cite{BD}) at the event
horizon of a black hole. The strong electric field which separates two 
oppositely 
charged particles beyond their Compton wavelength in the Schwinger
process is replaced by the event horizon in the case of Hawking
radiation. Further, the Hawking radiation was also explained as
an outcome of the so called Unruh effect in the relativistic picture (see \cite{BD}).
The argument of Unruh for such emission process is that an observer in an
accelerated frame will see radiation in the vacuum of inertial
observer (known as Unruh effect) \cite{UN1,UN2}. Whereas from inertial 
frame, there will be no radiation in
the vacuum states. Which therefore indicates that the vacuum is a relative 
concept. The Unruh effect predicts that an accelerating
observer will see black-body radiation in a true vacuum of an inertial 
observer. The temperature of the inertial vacuum as measured 
by the accelerated observer increases with the
magnitude of acceleration and is given by $T=T_U=\hbar \alpha/(2\pi c
k)$, known as the Unruh temperature. In other words, the
background appears to be warm from an accelerating reference frame.
The ground state for an inertial observer is seen as in thermodynamic
equilibrium with a non-zero temperature by the uniformly
accelerated observer. In presence of strong black hole gravitational
field near the event horizon, which is equivalent to an
accelerated frame without gravity, the temperature  of the vacuum will
be large enough to create particle and anti-particle pairs if $kT_U
> 2m_0c^2$, with $m_0$ the rest mass of the particle (anti-particle). 
However, all such explanations are associated with the relativistic 
quantum field theoretic approach of particle production.

In the present article we would like to show that when observed 
from a uniformly accelerated frame, then
in the non-relativistic approximation, where the concept of
quantum vacuum in the sense of particle-antiparticle creation and annihilation
does not exist, the particle production  process near
the event horizon of the black hole is more or less
like Fowler-Nordheim field emission \cite{FN} (see also \cite{AS}). 
In the non-relativistic picture, neither the Schwinger's mechanism nor 
the Unruh effect are applicable for the particle production process.

The main objective of this work is to draw some analogy with the electron emission process under the 
action of a strong electric field applied near the surface of a metal, with that of the creation of particles
by strong gravitational field at the vicinity of event horizon of a black hole. Since the Fowler-Nordheim equation
for cold field emission of electrons from the metal surface is non-relativistic one, we have made a non-relativistic
approximation of particle creation picture near the event horizon.

For the sake of completeness, we would like to present briefly the mechanism of cold field emission of electrons.
It is well known that
there are mainly three kinds of electron emission processes from
metal surface. These are (a) the thermal emission, (b) the photo emission and
(c) the cold emission or field emission. Among these processes, the
thermal emission can be explained classically. The only quantum
concept needed is the introduction of electron chemical potential
inside the metal. To explain photo-emission, the concept of old
quantum mechanics or quantum theory is sufficient. The cold emission or
field emission of electrons are the processes driven  by a strong
external electric field applied at the metal surface. This kind of emission can occur
even if the metal is at
extremely low temperature, i.e., the electron gas is strongly degenerate. This is the basic 
reason to call the field emission process 
also as cold emission. Unlike the thermal emission or photo-emission, 
the field emission can only be explained as the quantum
mechanical tunneling of electrons through surface barrier \cite{FN,AS}. 
It has no analogy with any classical process.
However, for the general type of surface driving potential, this purely
quantum mechanical problem can not be solved exactly. A
semi-classical approach, called WKB method is used to get tunneling
coefficient for general type surface barrier potential. 
Now to explain physically the mechanism of 
cold emission of electrons from the metal surface,  one may 
assume that because
of quantum fluctuation, electrons from the sea of degenerate electron
gas within the metal always try to tunnel out through the metallic
surface. 
The electrons which are just out side the metal surface because of 
fluctuation are like visible dolphins on the surface of a lake.
Now as an electron comes out, it induces an
image charge on the metal surface, which pulls it back and does not
allow the tunneled electrons to move far away from the metal surface in
the atomic scale. However, if some strong attractive
electrostatic field is applied near the metallic surface, then
depending on the magnitude of Fermi energy and the height of 
surface potential barrier, which is approximately equal to the 
work function of the metal, the electrons may overcome the effect of 
induced image charge on the metal surface and get liberated.
The field emission process was first theoretically explained by Fowler and Nordheim in \cite{FN} in their
Royal Society paper.

Now to compare the particle production process near the event horizon 
in the non-relativistic 
scenario, with that of Fowler-Nordheim field emission,
we consider the motion of a particle in a local rest frame in
presence of an uniform gravitational 
field. Which is equivalent to the uniformly
accelerated motion of the frame of reference in absence of gravity. 
We assume that the strong gravitational field 
produced by the black hole is almost uniform in local rest frame. 

In this article our intention is to  show that in the non-relativistic 
approximation, the creation of particles (anti-particles)
near the event horizon of the black hole is almost identical with the
Fowler-Nordheim field emission when observed from a
uniformly accelerated reference frame. To the best of our knowledge
such study has not been reported earlier. 

We have organized the
article in the following manner. In the next section 
we have developed a formalism to obtained the Schr$\ddot{\rm{o}}$dinger
equation of a particle (anti-particle) in presence of uniform 
gravitational field. An outline to obtain 
a solution of this equation has been
discussed in Appendix A. Whereas, a derivation to obtain single particle Finally we have given conclusion of our 
findings and discussed the future perspective of this work.
\section{Schr$\ddot{\rm{o}}$dinger Equation of a Particle Undergoing
Uniform Accelerated Motion}
Our study is based on the principle of equivalence, according to which a 
frame of reference undergoing an
accelerated motion in absence of gravitational field is equivalent to
a frame at rest in presence of a gravitational field.
To develop the quantum mechanical formalism for a particle undergoing
a uniform accelerated motion, we start with the single particle
classical Lagrangian in Rindler coordinates, 
which can be derived from the work in, e.g.,  \cite{MS,MAX,CG} (see also Appendix B of this article for a derivation
of the Lagrangian and Hamiltonian in Rindler space).
\begin{equation}
L=-m_0c^2\left [\left ( 1+\frac{\alpha x}{c^2}\right )^2 -\frac{v^2}{c^2}
\right ]^{1/2}
\end{equation}
where $\alpha$ is the constant acceleration in terms of which the proper acceleration of the Rindler frame is given
by $g=\alpha/(1+\alpha x/c^2)$, and is assumed to be along  
$x$-direction,
$v=u_x$, the particle velocity and $m_0$ is the rest mass of the
particle.
The three momentum vector of the particle can then be written as
\begin{equation}
\vec p=\frac{m_0\vec v}{\left [ \left (1+\frac{\alpha x}{c^2} \right )^2
-\frac{v^2}{c^2} \right ]^{1/2}}
\end{equation}
Hence the Hamiltonian of the particle  is given by
\begin{equation}
H=m_0c^2\left ( 1+\frac{\alpha x}{c^2}\right ) \left (
1+\frac{p^2}{m_0^2 c^2}\right )^{1/2}
\end{equation}
In the non-relativistic approximation with $m_0 c^2 \gg pc$, the above
Hamiltonian reduces to
\begin{eqnarray}
H&\approx& m_0c^2\left (1+\frac{\alpha x}{c^2}\right ) \left (
1+\frac{p^2}{2m_0^2c^2}\right )\nonumber \\
&=&
\left (1+\frac{\alpha x}{c^2}\right ) \left (
m_0c^2+\frac{p^2}{2m_0}\right )\nonumber \\
\end{eqnarray}
Now in the quantum mechanical picture, the classical dynamical
variables $x$, $\vec p$ and $H$ are treated as operators, with
the commutation relations
\begin{equation}
[x,p_x]=i\hbar~~{\rm{and}}~~ [x,p_y]= [x,p_z]=0     
\end{equation}
The Schr$\ddot{\rm{o}}$dinger equation for the particle is then given by
\begin{equation}
H\psi =\left ( 1+\frac{\alpha x}{c^2} \right ) \left ( m_0c^2
+\frac{p^2}{2m_0} \right )\psi =E\psi
\end{equation}
Using the representation
\[
p^2=-\frac{\hbar^2}{2m_0} \left (
\frac{\partial^2}{\partial x^2} +\frac{\partial^2}{\partial y^2}
+\frac{\partial^2}{\partial z^2} \right )
\]
we have after a little algebraic manipulation
\begin{eqnarray}
-\frac{\hbar^2}{2m_0} \left (
\frac{\partial^2}{\partial x^2} +\frac{\partial^2}{\partial y^2}
+\frac{\partial^2}{\partial z^2} \right ) \psi(x,y,z)&+& \frac{\alpha
Ex}{c^2}\psi \\ \nonumber  &=&E_k\psi
\end{eqnarray}
where the kinetic energy of the particle $E_k=E-m_0c^2$. It is quite
obvious that in the separable form, the solution of the above equation
may be written as \cite{FN}
\begin{equation}
\psi(x,y,z) =NX(x)\exp\left (-\frac{ip_yy}{\hbar}\right ) 
\exp\left (-\frac{ip_zz}{\hbar}\right )
\end{equation}
Substituting back in eqn.(7), we have
\begin{equation}
\frac{d^2 X}{d x^2}-\frac{2m_0 E\alpha}{\hbar^2 c^2}x
X(x) =-\frac{2m_0}{\hbar^2} \left ( E_k- \frac{p_\perp^2}{2m_0}
\right ) X(x)
\end{equation}
where
\[
\frac{p_\perp^2}{2m_0}= \frac{p_y^2+p_z^2}{2m_0}
\]
is the orthogonal part of kinetic energy. Hence the parallel
part of kinetic energy is given by
\[
E_{\vert\vert}=E_k-\frac{p_\perp^2}{2m_0}
\]
Let us put 
\[
\zeta=\left ( \frac{2m_0E\alpha}{\hbar^2c^2}\right )^{1/3}x
\]
a new dimensionless variable and 
\[
E^\prime=\frac{2m_0E_{\vert\vert}}{\hbar^2}\left (
\frac{\hbar^2c^2}{2m_0E\alpha}\right )^{2/3}
\]
as another dimensionless quantity.
Then it can very easily be shown that with $\xi=E^\prime-\zeta$, 
the above differential equation (eqn.(9)) reduces to
\begin{equation}
\frac{d^2 X}{d\xi^2}+\xi X=0
\end{equation}
This equation is of the same form as was obtained by Fowler and
Nordheim in their original work on field emission of electrons 
(see the equation before eqn.(7) in \cite{FN}). The identical mathematical
 structure of the differential equations results from the same kind of constant driving fields in both cases. 
In the case of Fowler-Nordheim emission, it is
the constant attractive electrostatic field derived from the potential 
of the form $C-Ex$,
where $C$ is the surface barrier, which is approximated with the work
function of the metal and $E$ is the uniform electrostatic field
near the metal surface. The quantity $C-Ex$ acts as the driving 
potential for cold emission. Whereas in the
case of black hole emission the driving force 
is the uniform gravitational field near the event horizon of
the black hole.

In Appendix A we outline how to obtain the solution of the 
differential equation given by eqn.(10). With this solution,
we have
\begin{eqnarray}
\psi(x,y,z)&=&N\exp\left (-i\frac{p_yy}{\hbar}\right )\exp\left (-
\frac{ip_zz}{\hbar}\right )\nonumber \\ &&
(E^\prime-\zeta)^{1/2}H_{1/3}^{(2)}\left [
\frac{2}{3}(E^\prime-\zeta)^{3/2}\right ]
\end{eqnarray}
where $N$ is the normalization constant.
Since we expect oscillatory solution also along $x$-direction in the 
asymptotic 
region, we have replaced $J_{1/3}(x)$ by
$H_{1/3}^{(2)}(x)$, the Hankel function of second kind. Now, from the previous 
definitions
\begin{eqnarray}
\xi=E^\prime -\zeta&=&\frac{2m_0E_{\mid\mid}}{\hbar^2} \left (
\frac{\hbar^2c^2}{2m_0E\alpha}\right )^{2/3}\nonumber \\ & -&\left ( \frac{2m_0E\alpha}
{\hbar^2/c^2} \right )^{1/3} x,\nonumber
\end{eqnarray}
if it is assumed that for some local rest frame at a distance $x_l$ from the centre of the black hole, 
in the asymptotic region, i.e., $x_l \gg$ the Schwarzschild  radius, the gravitational field
$\alpha=GM/x_l^2$, the  quantity $\xi$ as defined above can be
expressed in terms of $x_l$ in the following manner.
\[
\xi \sim ax_l^{4/3}-bx_l^{1/3}
\]
where $a$ and $b$ are real positive constants.
The argument of the Hankel function, which in the present
physical scenario is the 
appropriate solution for the differential equation, given by
eqn.(10), is large enough and positive in this asymptotic region. The
Hankel function can therefore be expressed as an oscillatory function \cite{AST} in this uniformly 
accelerated frame. This is to be noted that
here we are not talking about the variation of $\alpha$. It is a constant 
for a particular frame of reference, called local frame,  having
spatial coordinate $x_l$, or equivalently for a frame at rest in 
presence of an uniform gravitational field $\alpha$, known as local acceleration.
To make this point more transparent, we have considered a large 
number of uniformly accelerated frame of references
in the space out side a black hole, situated at a close 
proximity of event horizon to asymptotically far away from
the event horizon. Each of these frames are designated by the 
spatial coordinate $x_l$ in one dimension, measured from
the centre of the black hole. Here to keep one to one 
correspondence with Fowler-Nordheim field emission, we have assumed 
one dimensional configuration.

On the other hand if it is assumed that the uniform acceleration for
a local frame at $x_l$, close to the event horizon, is blue shifted, or in
other words the gravitational field is assumed to be blue shifted for
a local frame at rest at $x_l$ near the event horizon, one can write
\[
\alpha=\frac{GM}{x_l^2} \left [ 1-\frac{R_s}{x_l}\right ]^{-1/2}
\]
which gives the diverging value for $\alpha$ at the Schwarzs-child radius, 
i.e. for
$x_l=R_s=2GM/c^2$. Or in other words, if the uniformly accelerated frame is 
considered exactly at the event horizon. 
It is quite obvious that the value of $\xi$ is negative near the event
horizon and remain negative up to certain value of $x$ for the local 
rest frames for which $\alpha$'s are quite
large. To accommodate the negative values for $\xi$ for a set of 
local rest frames, we
make the following changes in the wave function in the negative
$\xi$ region.
We replace $\xi$ by $-\xi$, and then the modified form of
Hankel function is given by \cite{AST}
\[
H_{1/3}^{(2)}\left(\exp\left (\frac{3}{2}\pi i \right  )Q\right
)
\]
which may be expressed in terms of the modified Bessel function of first
kind and is given by
\[
-\frac{1}{\sin(\pi/3)}\left[ I_{-1/3}(Q)+\exp(i \pi/3)I_{1/3}(Q)
\right ]
\]
where $Q=2\xi^{3/2}/3$. 

Now  we define the particle density in the following manner in a 
particular local rest frame in
presence of gravitational field $\alpha$.
\[
n={\rm{constant}} \mid \psi \mid^2
\]
The number density
will be large enough for the local rest frames near the event horizon 
where $\xi$'s are
negative. This also follows from the expression for modified Bessel
function of first kind for large $Q$ as shown below 
\[
I_\nu(Q)\sim \frac{1}{(2\pi Q)^{1/2}} \exp(Q)
\]
The physical reason for large particle number density near the
event horizon is due to the strong gravitational field, which
produces more particles compared to far regions. This is also true in
the case of Fowler-Nordheim field emission. More strong the
electrostatic field more will be the electron emission rate.
Now it can very easily be shown that in this region 
the number density is given by
\[
n\sim \xi^{1/2} \exp(2Q)
\]
Of course the model is not valid exactly at the event
horizon. 

When $\xi$ becomes positive, which is true for a frame quite far away 
from the
event horizon, the wave function is given by the Hankel function. 

At $\xi=0$, although the Hankel function diverges, the wave function
vanishes in this particular frame of reference because of 
$\xi^{1/2}$ term. Same is true for the solution
for $\xi <0$, which matches exactly with $\xi>0$ solution at $\xi=0$.
Further the Hankel function asymptotically becomes oscillatory (exponential with imaginary argument) in
nature. The wave function for $\xi\longrightarrow \infty$ is given by
\[
\psi(\xi)\sim \xi^{-1/4}\exp\left [ -i\left (\xi -\frac{5\pi}{12}
\right ) \right ]
\]
Then  the particle density in some local rest frame at $x_l$, which is
far away from the event horizon, in presence of 
an uniform weak gravitational field is given by
\[
n(\xi \longrightarrow \infty) \sim (ax_l^{4/3}-bx_l^{1/3})^{-1/2}
\]

The value of
$\xi=0$ gives $x_l=(E_{\mid\mid}/E)(c^2/\alpha)$, the spatial coordinate 
of a local rest frame where the particle
density is exactly zero. If it is further
assumed that $E_{\mid\mid}=E$, then $x_l=c^2/\alpha$. Therefore the
coordinate point where $\xi$ switches over from negative value to
positive value, depends on the acceleration of the local frame.
Therefore we may divide the whole
space out side the black hole into effectively six regions: 
for the set of local rest frames in presence of 
uniform gravitational field, 
but far from the event horizon, 
the wave functions are oscillatory. For $\xi>0$
but not large enough, the wave functions can be expressed in those
frames in terms of
Hankel function of second kind. At $\xi=0$, the nature of the wave
functions from both $\xi \longrightarrow 0+$ and $\xi \longrightarrow
0-$ show that it should vanish. For $\xi <0$, but the magnitude is
not large enough, the wave functions can be expressed in terms of
modified Bessel function of first kind. Very close to the event
horizon, where $\xi$ is also less than zero but with very high in
magnitude, the number density shows exponential growth and asymptotically diverges. Finally nothing
can be said at and inside the event horizon.
\section{Conclusion}
In this work we have drawn some analogy of particle production near
the event horizon of a black hole with that of field 
emission or cold emission of electrons from
the metal surface in the non-relativistic scenario in a frame
undergoing uniform accelerated motion in an otherwise
flat space-time geometry. 
In the case of cold emission, the driving force is the strong external 
electrostatic field
applied near the metal surface. The strong electrostatic field helps 
the electrons to tunnel out through the surface
barrier. These electrons are liberated to the real world from the 
conduction band of the metal.
Further in the case of cold emission, only electrons are liberated.
Whereas for the black hole particle production,
it is the strong gravitational
field of the black hole near the event horizon the driving force, creating pairs. One particle of the pair goes
inside the black hole and the other one is emitted.
Further, in the case of black hole emission the pairs come out
from the quantum vacuum, where they are in the form of condensates, 
whereas electrons in the
conduction band are the constituents of degenerate Fermi gas. Therefore in
the non-relativistic approximation of black hole particle production,
the tunneling coefficient can not be
obtained following the formalism developed by Fowler and
Nordheim \cite{FN}. 

It is strongly believed that in the quantum field
theoretic approach in curved space-time, the creation of particles at
the event horizon is basically Schwinger type quantum tunneling
process. In this article we have shown
that in the non-relativistic approximation, it is also a tunneling
process, but may be identified as gravitational Fowler-Nordheim
emission.
Therefore in the non-relativistic scenario for black hole pair
creation, 
the formalism has to be developed considering particle
(anti-particle) which has already been tunneled out 
near the event horizon, i.e., outside the event horizon. Whereas in
the case of field emission, in the original work of Fowler and
Nordheim, the electrons are assumed to be free particles inside the
metal (free Fermi gas). 
\section{Appendix A}
Consider the differential equation
\begin{equation}
\frac{d^2X}{d\xi^2}+\xi X=0
\end{equation}
To get a solution, let us substitute $X(\xi)=\xi^n\psi(\xi)$, where $n$ is an unknown quantity. Then the above
differential equation reduces to
\begin{equation}
\xi^2 \frac{d^2 \psi}{d \xi^2} +2n \xi \frac{d\psi}{d\xi}+[n(n-1)+\xi^3]\psi=0
\end{equation}
Let $\xi=\beta z^{2/3}$, where $\beta$ is another unknown quantity. Then we have the reduced form of the above
equation as
\begin{equation}
z^2\frac{d^2 \psi}{dz^2}+\left ( n+\frac{1}{4}\right ) \frac{4}{3}z \frac{d \psi}{dz} +\frac{4}{9} [n(n-1)+
\beta^3z^2]\psi(z)=0
\end{equation}
Let us choose $n=1/2$, then we have
\begin{equation}
z^2\frac{d^2\psi}{dz^2}+z\frac{d\psi}{dz}+\left [ \frac{4}{9} \beta^3 z^2- \frac{1}{9}\right ]\psi(z)=0
\end{equation}
Finally choosing $\beta=(9/4)^{1/3}$, we get
\begin{equation}
z^2\frac{d^2\psi}{dz^2}+z\frac{d\psi}{dz} +\left ( z^2 -\frac{1}{9} \right )\psi(z)=0
\end{equation}
Comparing this differential equation with the standard form of Bessel 
equation
\begin{equation}
z^2\frac{d^2\psi}{dz^2}+z\frac{d\psi}{dz} +\left ( z^2 - \nu^2\right )\psi(z)=0
\end{equation}
whose solution is $J_\nu(z)$, Bessel function of order $\nu$
(Bessel function with negative order has no relevance)
or $H_\nu^{(2)}(z)$, the second kind Hankel function of order $\nu$.
Then depending on the physical situation, 
we have the  appropriate solution of eqn.(17) as
\begin{equation}
\psi(z)=J_{1/3}(z) ~~{\rm{or}}~~ \psi(z)=H_{1/3}^{(2)}(z)
\end{equation}
\section{Appendix B}
In this Appendix using some of the established useful formulas of
special relativity with uniform accelerated motion (see \cite{MS,MAX,CG}) we shall
obtain the single particle Lagrangian and
Hamiltonian in Rindler space.
Using the results from \cite{MS,MAX,CG}) the Rindler coordinates are given by
\begin{eqnarray}
ct&=&\left (\frac{c^2}{\alpha}+x^\prime\right )\sinh\left (\frac{\alpha t^\prime}
{c}\right ) ~~{\rm{and}}~~ \nonumber \\
x&=&\left (\frac{c^2}{\alpha}+x^\prime\right )\cosh\left (\frac{\alpha t^\prime}
{c}\right ) 
\end{eqnarray}
Hence one can also express the inverse relations
\begin{equation}
ct^\prime=\frac{c^2}{2\alpha }\ln\left (\frac{x+ct}{x-ct}\right )
~~{\rm{and}}~~ x^\prime=(x^2-(ct)^2)^{1/2}-\frac{c^2}{\alpha }
\end{equation}
The Rindler space-time coordinates, given by eqns.(19) and (20) 
are then just an accelerated frame
transformation of the Minkowski metric of special relativity. The
Rindler coordinate transform the Minkowski line element
\begin{eqnarray}
ds^2&=&d(ct)^2-dx^2-dy^2-dz^2 ~~{\rm{to}}~~\nonumber \\ ds^2&=&\left
(1+\frac{\alpha x^\prime}{c^2}\right)^2d(ct^\prime)^2-{dx^\prime}^2
-{dy^\prime}^2-{dz^\prime}^2
\end{eqnarray}
The general form of 
metric tensor may then be written as
\begin{equation}
g^{\mu\nu}={\rm{diag}}\left (\left (1+\frac{\alpha x}{c^2}\right
)^2,-1,-1,-1\right )
\end{equation}
Now following the concept of relativistic dynamics of special theory
of relativity \cite{LL}, the action
integral may be written as (see also \cite{CG})
\begin{equation}
S=-\alpha_0 \int_a^b ds\equiv \int_a^b Ldt
\end{equation}
Then using eqns.(19)-(22) and putting $\alpha_0=-m_0 c$, where $m_0$ is the
rest mass of the particle, the Lagrangian of the particle is given by
\begin{equation}
L=-m_0c^2\left [\left ( 1+\frac{\alpha x}{c^2}\right )^2 -\frac{v^2}{c^2}
\right ]
\end{equation}
where $\vec v$ is the three velocity of the particle. The three momentum of the
particle is then given by
\[
\vec p=\frac{\partial L}{\partial \vec v}, ~~ {\rm{or}}
\]
\begin{equation}
\vec p=\frac{m_0\vec v}{\left [ \left (1+\frac{\alpha x}{c^2} \right )^2
-\frac{v^2}{c^2} \right ]^{1/2}}
\end{equation}
Hence the Hamiltonian of the particle is given by
\[
H=\vec p.\vec v-L ~~ {\rm{or}}
\]
\begin{equation}
H=m_0c^2 \left (1+\frac{\alpha x}{c^2}\right ) \left (1+
\frac{p^2}{m_0^2c^2}\right )^{1/2}
\end{equation}
which is eqn.(3) in the main text.

\noindent Acknowledgment: We are extremely grateful to the anonymous referee for constructive criticism and helping us in
making the paper more beautiful.

\end{document}